\documentclass[pra,twocolumn,showpacs,nopreprintnumbers,amsmath,amssymb]{revtex4}
\usepackage{graphicx}
\usepackage{dcolumn}
\usepackage{bm}

\begin{document}

\preprint{}

\title{
Initial wave packets and the various power-law decreases of \\
scattered wave packets at long times
}

\author{Manabu Miyamoto}%
 \email{miyamo@hep.phys.waseda.ac.jp }
\affiliation{%
Department of Physics, Waseda University, 3-4-1 Okubo, Shinjuku-ku, 
Tokyo 169-8555,  Japan 
}%

\date{\today}

\begin{abstract} 
The long time behavior of scattered wave packets $\psi (x,t)$ from 
a finite-range potential is investigated, 
by assuming $\psi (x,t)$ to be initially located outside the potential. 
It is then shown that $\psi (x,t)$ can asymptotically decrease 
in the various power laws at long time, 
according to its initial characteristics at small momentum. 
As an application, we consider the square-barrier potential system 
and demonstrate that $\psi (x,t)$ exhibits the asymptotic behavior $t^{-3/2}$, 
while another behavior like $t^{-5/2}$ can also appear for another $\psi (x,t)$. 
\end{abstract}

\maketitle


A long time deviation from the exponential decay law is predicted 
in unstable quantum systems such as an $\alpha$-decaying nucleus 
\cite{Khalfin}. 
These systems are often modeled in 
the systems of a particle in a short range potential. 
In this approach, 
a nonexponential decay is generally found to follow a particular power law, $t^{-3/2}$ 
(see, e.g., Refs.\ \cite{Rauch,Muga(1995),Dijk(2002),Garcia(1995)}). 
On the contrary to the theoretical results, 
an experimental evidence still has not been discovered \cite{Greenland}, 
which requires us 
to reexamine such a nonexponentially decaying behavior from every aspect. 
As far as the author knows, conventional studies are developed 
without taking into account of the characteristics of the initial wave packets. 
A consideration for them may lead to a discovery of other aspects of the subjects. 
Indeed, as for the free particle system, 
the various power decreases of wave packets, not restricted to $t^{-3/2}$, 
have recently been demonstrated and shown to be characterized 
by the behavior of initial wave packets at small momentum 
\cite{Unnikrishnan,Lillo,Damborenea,Miyamoto}. 

In this paper, we analytically prove that 
the wave packet $\psi (x,t)$ scattered from a finite-range potential 
can asymptotically behave like $t^{-j/2}$ at long times, 
where $j=1, 2, \ldots.$ 
By making an assumption that the $\psi (x,t)$ is initially located outside the potential, 
these various power behaviors can be characterized by 
the initial characteristics of the $\psi (x,t)$ 
at small momentum, like for the free particle case \cite{Damborenea,Miyamoto}. 
The validity of our analysis is numerically confirmed, 
through the application to the system with a square barrier potential.

For a one-dimensional system with a potential $V(x)$, 
the Hamiltonian $H$ is defined as $H\equiv H_0 + V$, 
where $H_0 \equiv  -(\hbar^2 /2M) d^2/dx^2$ being the free Hamiltonian. 
For simplicity, we use the unit such that $\hbar =1$ and $2M=1$ throughout the paper. 
Although we confine ourselves to the one dimensional case, the following discussion 
can be extended to that in an arbitrary dimension in principle. 
The potential $V(x)$ is assumed to have a finite-range, i.e., 
\begin{equation}
V(x) =0 ~~~\mbox{ for } |x|>R, 
\label{eqn:10}
\end{equation}
for a positive number $R$. 
We only consider the systems without bound states. 
This restriction however is easily moderated. 
By using the eigenfunction expansion, 
the wave packet $\psi (x,t)$, 
evolving from the initial state (wave packet) $ \psi (x)$, 
is expressed as 
\begin{equation}
\psi (x,t) = (e^{-itH} \psi)(x) = 
\int_{-\infty}^{\infty} 
e^{-itk^2} \varphi(x,k) 
\widetilde{\psi} (k) dk, 
\label{eqn:20}
\end{equation}
where the $\widetilde{\psi} (k)$, 
determining the initial energy-distribution, is defined by 
\begin{equation}
\widetilde{\psi} (k) \equiv \int_{-\infty}^{\infty} 
\varphi^* (y,k) \psi (y) dy . 
\label{eqn:25}
\end{equation}
The functions $\varphi(x,k)$ ($k \in \mathbb{R}\backslash \{ 0\}$) 
are stationary scattering solutions of 
the time-independent Schr{\" o}dinger equation,   
\begin{equation}
       [H_0+V] \varphi (x,k) = k^2 \varphi(x,k) . 
       \label{eqn:30}
\end{equation}
More precisely, 
they must satisfy the Lippmann-Schwinger equations in one dimension \cite{Muga(2001)}, 
\begin{equation}
       \varphi (x,k) = 
       \frac{e^{i kx}}{\sqrt{2\pi}}
       \mp \frac{1}{2i |k|} 
                    \int_{-\infty}^{\infty} 
                    e^{ \mp i |k| |x-y|} V(y) ~
                    \varphi(y,k) ~d y  .
       \label{eqn:40}
\end{equation}
Either solution of the above equation with ($+$) or ($-$) sign 
is accepted for $\varphi(x,k)$ in Eq.\ (\ref{eqn:20}). 
Notice that, being integral equations, 
the Lippmann-Schwinger equations already incorporate the boundary conditions, 
unlike Eq.\ (\ref{eqn:30}). 
We here choose the equation with ($+$) sign. 
Then, $\varphi (x,k)$ for positive $k$ and for negative $k$ 
are solutions of Eq.\ (\ref{eqn:30}), 
to the cases of an incident plane wave from the left 
and from the right of the potential, respectively.

We first derive an asymptotic expansion of $\psi (x,t)$ 
at long times. 
The integral in Eq.\ (\ref{eqn:20}) is changed to  
a Fourier-integral form over the energy variable $E=k^2$, i.e., 
\begin{equation}
\psi(x,t) 
=
\frac{1}{2} \sum_{\sigma=\pm} \int_{0}^{\infty} 
E^{-1/2} {\cal F}_{\sigma} (x,E) e^{-itE} dE, 
\label{eqn:n10}
\end{equation}
where ${\cal F}_{\pm} (x,E) \equiv  \varphi(x,\pm E^{1/2}) 
\widetilde{\psi}(\pm E^{1/2} ) $. 
We may decompose ${\cal F}_{\pm} (x,E)$ into the following forms, 
\begin{equation}
{\cal F}_{\pm} (x,E) =  \pm E^{1/2} {\cal O}_{\pm} (x,E) + {\cal E}_{\pm} (x,E), 
\label{eqn:n25}
\end{equation}
where 
\begin{equation}
{\cal O}_{\pm} (x,E)=\sum_{r+s={\rm odd}}^{\infty} \frac{E^{(r+s-1)/2}}{r!s!}
\partial_k^{r} \varphi (x,\pm 0)  \widetilde{\psi}^{(s)} (\pm 0), 
\label{eqn:n30}
\end{equation}
\begin{equation}
{\cal E}_{\pm} (x,E)=\sum_{r+s={\rm even}}^{\infty} \frac{E^{(r+s)/2}}{r!s!}
\partial_k^{r} \varphi (x,\pm 0)  \widetilde{\psi}^{(s)} (\pm 0), 
\label{eqn:n40}
\end{equation}
as $E \rightarrow 0$. Both $\varphi (x,k)$ and $\widetilde{\psi} (k)$ 
are assumed to be differentiable with respect to $k$ without the origin. 
Furthermore, we have used the notations that 
$\partial_k^{n} \varphi (x,\pm 0) 
\equiv \lim_{k \rightarrow \pm 0} \partial^{n} \varphi (x,k)/\partial k^{n}$, 
and for any function of $k$, say, $f(k)$, 
$f^{(n)} (\pm 0) \equiv \lim_{k \rightarrow \pm 0} d^{n} f(k) / dk^{n}$. 
Note that by setting Eq.\ (\ref{eqn:n25}) into (\ref{eqn:n10}), 
${\cal O}_{\pm} (x,E)$ makes no singularity at $E=0$ while 
$E^{-1/2} {\cal E}_{\pm} (x,E)$ does not necessarily.  
Then, in taking such a singularity into account, 
the asymptotic form of the Fourier integral (\ref{eqn:n10}) may read formally 
\cite{Copson} 
\begin{equation}
\begin{array}{rcl}
\psi(x,t) 
&\sim&
\displaystyle{
\frac{1}{2} \sum_{j=0}^{\infty }  
\frac{1}{(it)^{j+1}} 
\sum_{\sigma = \pm } \sigma \partial_E^j {\cal O}_{\sigma} (x,0) 
}\\
&&
\displaystyle{
+ \frac{1}{2} \sum_{j=0}^{\infty }  
\frac{\Gamma (j+1/2) }{j! (it)^{j+1/2}} 
\sum_{\sigma = \pm } \partial_E^j {\cal E}_{\sigma} (x,0), 
}
\end{array}
\label{eqn:n50}
\end{equation}
as $t\rightarrow \infty$, 
where 
$\lim_{E \rightarrow \infty} \partial_E^j {\cal O}_{\pm} (x,E) =0$ and 
$\lim_{E \rightarrow \infty} \partial_E^j {\cal E}_{\pm} (x,E) =0$ 
were assumed. 
It is noted that 
we could also obtain the expansion (\ref{eqn:n50}) by different methods 
(see, Refs.\ 
\cite{Rauch,Muga(1995),Dijk(2002),Garcia(1995)} and references therein).

As same as the results in Refs. 
\cite{Unnikrishnan,Lillo,Damborenea,Miyamoto,Nakazato}, 
we can expect from Eq.\ (\ref{eqn:n50}) that 
by specifying the low-energy behaviors of 
both ${\cal O}_{\pm} (x,E)$ and ${\cal E}_{\pm} (x,E)$ appropriately, 
$\psi (x,t)$ can asymptotically show the various power-decreases. 
This situation may be realized by 
such an initial wave packet $\psi (x)$ that satisfies 
\begin{equation}
\widetilde{\psi}^{(n)}(\pm 0) =0 ~~~\mbox{ for } n=0, 1, \ldots, m-1, 
\label{eqn:n60}
\end{equation}
with a certain integer $m$. 
In fact, let us first consider the case of $m$ 
being an even number given by $m=2\overline{m}$. 
In this case, ${\cal O}_{\pm} (x,E)$ and ${\cal E}_{\pm} (x,E)$ read 
\begin{equation}
{\cal O}_{\pm} (x,E)=
\frac{E^{m/2}}{m!} \partial_k \varphi (x,\pm 0) \widetilde{\psi}^{(m)} (\pm 0)
+O(E^{(m+2)/2}), 
\label{eqn:n70}
\end{equation}
\begin{equation}
{\cal E}_{\pm} (x,E)=
\frac{E^{m/2}}{m!} \varphi (x,\pm 0) \widetilde{\psi}^{(m)} (\pm 0)
+O(E^{(m+2)/2}), 
\label{eqn:n80}
\end{equation}
as $E \rightarrow 0$, respectively. 
Then, substituting them into Eq.\ (\ref{eqn:n50}) leads to an expected result that 
\begin{equation}
\begin{array}{rcl}
\psi(x,t) 
&\sim&
\displaystyle{
\frac{1}{2} \sum_{\sigma = \pm } \Biggl[ 
\sigma \frac{1}{(it)^{\overline{m}+1}} 
\partial_E^{\overline{m}} {\cal O}_{\sigma} (x,0) 
}\\
&&
\displaystyle{
+
\frac{\Gamma (\overline{m}+1/2) }{\overline{m}! (it)^{\overline{m}+1/2}} 
\partial_E^{\overline{m}} {\cal E}_{\sigma} (x,0)
\Biggr] + O(t^{-\overline{m}-3/2}), 
}
\end{array}
\label{eqn:n90}
\end{equation}
where 
$\partial_E^{\overline{m}} {\cal O}_{\pm} (x,0) 
=\overline{m}! \partial_k \varphi (x,\pm 0) \widetilde{\psi}^{(m)} (\pm 0)/m!$ 
and $\partial_E^{\overline{m}} {\cal E}_{\pm} (x,0) 
=\overline{m}! \varphi (x,\pm 0) \widetilde{\psi}^{(m)} (\pm 0)/m!$. 
On the other hand, 
if $m$ is an odd integer $m=2\overline{m}-1$, one see that 
\begin{equation}
{\cal O}_{\pm} (x,E)=
\frac{E^{(m-1)/2}}{m!} \varphi (x,\pm 0) \widetilde{\psi}^{(m)} (\pm 0)
+O(E^{(m+1)/2}), 
\label{eqn:n100}
\end{equation}
\begin{equation}
\begin{array}{rcl}
{\cal E}_{\pm} (x,E)
&=& 
\displaystyle{
\frac{E^{(m+1)/2}}{(m+1)!} 
\Bigl[ 
(m+1) \partial_k \varphi (x,\pm 0) \widetilde{\psi}^{(m)} (\pm 0) 
}
\\
&&
\displaystyle{
+\varphi (x,\pm 0) \widetilde{\psi}^{(m+1)} (\pm 0)
\Bigr]
+O(E^{(m+3)/2}), 
}
\end{array}
\label{eqn:n110}
\end{equation}
as $E \rightarrow 0$. Inserting them into Eq.\ (\ref{eqn:n50}) again, we obtain 
\begin{equation}
\begin{array}{rcl}
\psi(x,t) 
&\sim&
\displaystyle{
\frac{1}{2} \sum_{\sigma = \pm } \Biggl[ 
\sigma \frac{1}{(it)^{\overline{m}}} 
\partial_E^{\overline{m}-1} {\cal O}_{\sigma} (x,0) 
}\\
&&
\displaystyle{
+
\frac{\Gamma (\overline{m}+1/2) }{\overline{m}! (it)^{\overline{m}+1/2}} 
\partial_E^{\overline{m}} {\cal E}_{\sigma} (x,0)
\Biggr] + O(t^{-\overline{m}-1}), 
}
\end{array}
\label{eqn:n120}
\end{equation}
where 
$\partial_E^{\overline{m}-1} {\cal O}_{\pm} (x,0) 
=(\overline{m}-1)! \varphi (x,\pm 0) \widetilde{\psi}^{(m)} (\pm 0)/m!$ 
and $\partial_E^{\overline{m}} {\cal E}_{\pm} (x,0) 
=\overline{m}! [(m+1) \partial_k \varphi (x,\pm 0) \widetilde{\psi}^{(m)} (\pm 0) 
+\varphi (x,\pm 0) \widetilde{\psi}^{(m+1)} (\pm 0) ]/(m+1)!$.

We have to give such initial wave packets $\psi (x)$ 
satisfying the formal condition (\ref{eqn:n60}). 
Then, it seems practically advantageous to rewrite 
this condition 
in terms of the initial wave packet $\widehat{\psi} (k)$ 
in momentum (or $H_0$) representation. 
This can be achieved by assuming the $\psi (x)$ 
to be located in the left of the scattering potential $V(x)$, i.e., 
\begin{equation} 
\psi (x) =0 ~~~\mbox{ for }  x \geq -R .
\label{eqn:110}
\end{equation}
This assumption will however be relaxed to that $\psi (x) =0$ for $|x| \leq R$, 
in the discussions below. 
From the assumption (\ref{eqn:110}), 
the $\widehat{\psi} (k)$ is expressed 
by the integral over the truncated interval $(-\infty,-R)$,
\begin{equation}
\widehat{\psi} (k) 
= \int_{-\infty}^{\infty} \frac{e^{-iky}}{\sqrt{2\pi}} \psi (y) dy 
= \int_{-\infty}^{-R} \frac{e^{-iky}}{\sqrt{2\pi}} \psi (y) dy . 
\label{eqn:140}
\end{equation}
Meanwhile, the assumption (\ref{eqn:10}) implies that $\varphi (x,k)$ for $x<-R$ 
is written by a superposition of plane waves  
\begin{equation}
\varphi (x,k) = 
[ g_{+} (k) e^{i|k|x} + g_{-} (k) e^{-i|k|x} ]/\sqrt{2\pi} . 
\label{eqn:120} 
\end{equation}
Since we adopt Eq.\ (\ref{eqn:40}) with ($+$) sign, 
we see that $g_{+} (k) =1$ for $k >0$ and $0$ for $k <0 $. 
Substituting Eqs.\ (\ref{eqn:140}) and (\ref{eqn:120}) 
into (\ref{eqn:25}), 
we can obtain a desirable expression for $\widetilde{\psi}^{(n)} (\sigma 0)$ 
as a linear combination of $\widehat{\psi}^{(l)} (\pm 0)$'s, 
\begin{eqnarray}
\hspace*{-7mm} \widetilde{\psi}^{(n)} (\sigma 0) 
\hspace*{-1mm} &=& \hspace*{-1mm} 
(\sigma 1)^{n} \delta_{+, \sigma} 
\ \widehat{\psi}^{(n)} (+0)  
\nonumber \\
\hspace*{-1mm} && \hspace*{-1mm} 
+ \sum_{l=0}^{n} {{n}\choose{l}} 
 (-\sigma 1)^{l} {g_{-}^*}^{(n-l)} (\sigma 0)  
\ \widehat{\psi}^{(l)} (-0) , 
\label{eqn:130}
\end{eqnarray}
where $\sigma0$ stands for the limit symbol $+0$ or $-0$ for $\sigma=+$ or $-$, 
respectively, 
and $\delta_{+, \sigma}$ denotes Kronecker's delta. 
From this, 
one might expect that 
the condition (\ref{eqn:n60}) at low energy, 
which determines the asymptotic forms (\ref{eqn:n90}) and (\ref{eqn:n120}), 
is characterized by the condition at small momentum \cite{example}, 
\begin{equation}
\widehat{\psi}^{(l)} (\pm 0) =0~~~\mbox{ for } l=0, 1, \ldots, m-1. 
\label{eqn:240}
\end{equation}
This condition indeed implies 
the condition (\ref{eqn:n60}). However the converse does not necessarily hold. 
This incompatibility comes from the actual 
behavior of $g_{-}^{(n-l)} (\pm 0)$'s.

Let us now demonstrate the asymptotic formulas (\ref{eqn:n90}) and (\ref{eqn:n120}) 
with Eq.\ (\ref{eqn:130}), 
by applying it to 
the system with the square barrier potential $V(x)$ given by Eq.\ (\ref{eqn:10}) 
and 
\begin{equation} 
V(x) =  V_0   ~~~\mbox{ for } |x|\leq R , 
\label{eqn:260}
\end{equation}
where $V_0$ ($>0$) 
is the height of the potential barrier. 
For this system, $g_{-}(k)$ in Eq.\ (\ref{eqn:120}) 
is given by (see, e.g., \cite{Schiff})
\begin{equation}
g_{-}(k)= \left\{ 
                \begin{array}{ll}
                      \displaystyle{ 
                      \frac{\rho^2 + k^2}{2i k \rho } 
                      g(k) \sinh 2\rho R }
                      &\mbox{ for }  0<k<k_b \\ 
                      g(-k)  & \mbox{ for } -k_b <k<0  
                \end{array}
        \right.  ,
\label{eqn:280}
\end{equation}
where $k_b ={V_0}^{1/2}$, $\rho = [{k_b}^2 -k^2 ]^{1/2}$, and 
\begin{equation}
                 g(k)\equiv \biggl[ 
                  \cosh 2\rho R + 
                  \frac{k^2 - \rho^2 }{2ik \rho } 
                  \sinh 2\rho R 
                  \biggr]^{-1} e^{-2i kR}  .
\label{eqn:290}
\end{equation}
The $g(k)$ has the remarkable properties:  
$\lim_{k \rightarrow \pm 0} g(k) =0 $ and 
$\lim_{k \rightarrow \sigma 0} g(\sigma k)/k 
= \sigma 2/(ik_b  \sinh 2k_b R ) $, 
to give 
\begin{equation}
g_{-}(+0) =-1 ,~~~ 
g_{-}(-0) =0 . 
\label{eqn:310}
\end{equation}
This implies that incident plane waves 
with vanishing energy are totally reflected by the barrier. 
Since the $\varphi (x, k)$ is continuous even at $x=\pm R$, we consequently see that 
\begin{equation}
\varphi (x, \pm 0)=0   ~~~\mbox{ for }x \in \mathbb{R} .
\label{eqn:312}
\end{equation}
This is just the case without the zero-energy resonance \cite{Rauch,Bianchi}. 
It is worth noting that if we assume that 
the initial wave packet $\psi (x)$ is absolutely integrable, 
Lebesgue's dominated convergence theorem leads to 
\begin{equation}
\widetilde{\psi}(\pm 0) = \int_{-\infty}^{\infty} 
\varphi^* (y,\pm 0) \psi (y) dy =0 . 
\label{eqn:317}
\end{equation}
Hence the condition (\ref{eqn:n60}) for $m=1$ can be always satisfied 
for this system. 
This result implies the use of Eq.\ (\ref{eqn:n120}) with $\overline{m}=1$, 
which leads to an estimation 
such that $\psi (x,t)$ behaves as $t^{-3/2}$ \cite{Rauch} 
, i.e.,   
\begin{equation}
\psi (x,t) \sim 
\frac{1}{2} 
\frac{\Gamma (3/2) }{(it)^{3/2}} 
\sum_{\sigma = \pm } 
\partial_k \varphi (x,\sigma 0) \ \widetilde{\psi}^{(1)} (\sigma 0) .
\label{eqn:350}
\end{equation}
However, the determination of the asymptotic behavior of $\psi (x,t)$ may also need 
a consideration to the concrete behavior of the $\psi (x)$. 
For definiteness, we here confine ourselves to deriving the behavior like $t^{-5/2}$, 
and assume that the $\psi (x)$ satisfies Eq.\ (\ref{eqn:110}) and 
has a continuity at zero momentum: 
\begin{equation}
\widehat{\psi}^{(n)}(+0)=\widehat{\psi}^{(n)}(-0) 
~ \mbox{ for } n=0, 1, 2, 3. 
\label{eqn:315}
\end{equation}
This allows us to represent both $\widehat{\psi}^{(n)}(+0) $ and 
$\widehat{\psi}^{(n)}(-0) $ by the same symbol $\widehat{\psi}^{(n)}(0)$. 
In this case, 
substitution of Eqs.\ (\ref{eqn:310}) and (\ref{eqn:317}) into (\ref{eqn:130}) 
leads to a simple expression for $\widetilde{\psi}^{(1)} (\pm 0)$: 
\begin{equation}
\widetilde{\psi}^{(1)} (\sigma 0) 
=
{g_{-}^*}^{(1)} (\sigma 0)  \ 
\widehat{\psi} (0)  
+
2\delta_{+, \sigma} \widehat{\psi}^{(1)} (0)  , 
\label{eqn:370}
\end{equation}
where $g_{-}^{(1)} (\pm 0)$ are straightforwardly 
evaluated as 
\begin{eqnarray}
&g_{-}^{(1)} (+0)=2iR + 2(\cosh 2k_b R) /(ik_b \sinh 2k_b R) , & \\ 
&g_{-}^{(1)} (-0)=-2/(ik_b \sinh 2k_b R) . &
\end{eqnarray}
Equation (\ref{eqn:370}) implies the fact that 
if $\psi (x)$ satisfies the condition (\ref{eqn:240}) for $m=2$, i.e., 
\begin{equation}
\widehat{\psi} (0) =0 ~~~ \mbox{ and } ~~~ \widehat{\psi}^{(1)} (0) =0 , 
\label{eqn:360}
\end{equation}
then, $\widetilde{\psi}^{(1)} (\pm 0) =0$ and 
the formula (\ref{eqn:350}) is no longer effective. 
Note that this does not immediately mean 
the case of $m=2$ in the condition (\ref{eqn:n60}) realized. 
Because, as is pointed out after Eq.\ (\ref{eqn:240}), 
the system's properties (\ref{eqn:310}) with Eqs.\ (\ref{eqn:315}) and 
(\ref{eqn:360}) causes $\widetilde{\psi}^{(2)} (\pm 0) =0 $, 
even if $\widehat{\psi}^{(2)} (0) \neq 0$. See Eq.\ (\ref{eqn:130}). 
Therefore, we should actually refer to the case of $m=3$ 
in the condition (\ref{eqn:n60}). 
This time, the formula (\ref{eqn:n120}) (with $\overline{m}=2$) is used again to read 
\begin{equation}
\psi (x,t) \sim 
\frac{1}{2} 
\frac{\Gamma (5/2) }{6(it)^{5/2}} 
\sum_{\sigma = \pm } 
\partial_k \varphi (x,\sigma 0) \ \widetilde{\psi}^{(3)} (\sigma 0) . 
\label{eqn:380}
\end{equation}
In this case, 
$
\widetilde{\psi}^{(3)} (\sigma 0) 
=
3 {g_{-}^*}^{(1)} (\sigma 0)  \ 
\widehat{\psi}^{(2)} (0)  
+
2\delta_{+, \sigma} \widehat{\psi}^{(3)} (0) 
$.

In order to illustrate the above analysis, we consider 
the long-time behavior of the nonescape probability $P(t)$ 
\begin{equation}
P(t)\equiv \int_{a}^{b} |\psi (x,t)|^2 dx, 
\label{eqn:6}
\end{equation}
instead of $\psi(x,t)$ itself. This is the probability of finding 
a particle, initially prepared in state $\psi$, 
in a bounded interval $I=[a,b]$ at a later time $t$. 
By substituting Eq.\ (\ref{eqn:n90}) or (\ref{eqn:n120}) 
into the definition (\ref{eqn:6}), 
the asymptotic behavior of $P(t)$ directly reflects that of $\psi (x,t)$. 
We here restrict ourselves to 
the three wave packets $\phi_0 (x)$, $\phi_1 (x)$, and $\phi_2 (x)$, 
as the initial wave packets $\psi (x)$: 
\begin{equation}
\widehat{\phi}_{m} (k) = N_{m} k^{m} e^{-{a_0}^2 (k-k_0 )^2 /2 -ikx_0}, ~~~
\label{eqn:320}
\end{equation}
where $m=0, 1, 2$, $a_0 >0$, $k_0, x_0 \in \mathbb{R}$, and 
$N_{m}$ 
being the normalization constants. 
Note that the parameters $a_0$ and $x_0$ roughly indicate 
the width and the location of $\phi_{m} (x)$, respectively. 
These wave packets are rapidly-decreasing functions and satisfy 
the regularity (\ref{eqn:315}). 
Both $\phi_0 (x)$ and $\phi_1 (x)$ are chosen 
to confirm the asymptotic formula (\ref{eqn:350}), 
while $\phi_2 (x)$ satisfies the assumption (\ref{eqn:360}) 
to realize the asymptotic formula (\ref{eqn:380}). 
These wave packets do not satisfy the assumption (\ref{eqn:110}). 
This situation however could be taken into account. 
In fact, if we take appropriate parameters $a_0$ and $x_0$ 
where the latter satisfies $|x_0|=-x_0 >> R$, 
the errors in $\widetilde{\phi}_{m}^{(n)} (\pm 0)$'s 
might be negligibly small \cite{errors}. 
\begin{figure}
\rotatebox{270}{
\includegraphics[width=0.3\textwidth]{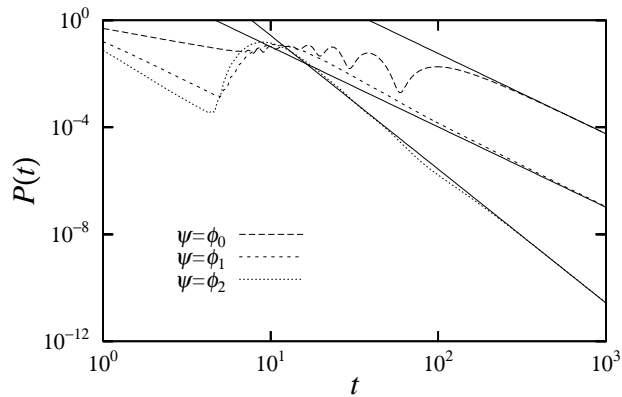}
}
\caption{\label{fig:figure1} 
Nonescape probabilities $P(t)$'s for initial wave packets $\psi=$ 
$\phi_0$, $\phi_1$, and $\phi_2$  
and their asymptotes predicted by 
Eq.\ (\ref{eqn:350}) or (\ref{eqn:380}) (solid lines). 
For $\phi_0$ and $\phi_1$,  
$P(t)$ shows the well-known $t^{-3}$ behavior at long times 
(long-dashed and short-dashed lines), 
whereas $P(t)$ for $\phi_2$ exhibits 
another power-law behavior like $t^{-5}$ (a dotted-line). 
}
\end{figure}

Figure \ref{fig:figure1} 
shows the time evolution of $P(t)$, 
and the asymptotes predicted by 
Eq.\ (\ref{eqn:350}) or (\ref{eqn:380}), 
for the three initial wave packets $\phi_0 (x)$, $\phi_1 (x)$, and $\phi_2 (x)$. 
In our calculation, we have chosen a set of parameters 
$a_0 =1.0$, $k_0 =1.0$, and $x_0=-20.0$ for the three initial wave packets. 
Then, every average momentum in these initial states 
is positive. 
We have also chosen in all these cases 
the potential range $R=1.0$, which is much smaller than the location $|x_0|=20.0$, 
and height $V_0=16.0$, which is greater than the expectation value of energy 
$\langle \phi_{m} , H_0 \phi_{m} \rangle$. 
The interval $I=[a, b]$ for the nonescape probability is set 
around the initial location of the wave packet $x_0$, and we set $a=-22.0$ and $b=-18.0$. 
One may recognize three different regions in the figure: 
for small $t$, all $P(t)$'s decrease smoothly, and then, 
they partially revive before decreasing again. 
These regions reflect the motion of 
a wave packet, i.e., it leaves the interval $I$ for the barrier, 
returns to the interval after a collision with the barrier, 
and goes outside through the interval, respectively \cite{situation}. 
It is clearly seen 
that, in the last region, 
$P(t)$'s for initial wave packets $\phi_0$ and $\phi_1$ 
approach to the asymptote 
parallel to the well-known $t^{-3}$. 
On the other hand, 
the behavior of $P(t)$ for the initial state $\phi_2$ is in quite agreement with 
the asymptote parallel to $t^{-5}$, other than $t^{-3}$.

To summarize, 
we have considered the finite-range potential-systems for one dimension 
and explicitly characterized the various power decreasing behaviors 
of the scattered wave-packets at long times, 
in terms of their position and momentum bahavior at an initial time. 
Our results can also cover the free case of 
Refs.\ \cite{Damborenea,Miyamoto} and that of Refs. \cite{Unnikrishnan,Lillo} 
with a slight modification. 
The power-law decrease of the potential systems at long times 
still has not been observed experimentally, 
however it may exhibit an interesting phenomenon, involving a peculiar structure 
where the characteristics of the initial state play a crucial role. 
%


The author would like to thank Professor I.\ Ohba 
and Professor H.\ Nakazato for useful and helpful discussions.



\begin{thebibliography}{99}
\bibitem{example} 
We can find the wave packets that 
satisfy the assumptions (\ref{eqn:110}) and (\ref{eqn:240}) 
simultaneously, 
in $C_0^{\infty}$-functions, e.g., 
$(m-1)$-th derivative of the function defined by 
$\psi (x) = \exp [-a_0^2 /[d^2 -(x-x_0)^2 ]]$ for $|x-x_0 |<d$ or 
$0$ otherwise, where $a_0 >0$, $d>0$, $x_0 <0$, and $x_0 +d < -R$.
\bibitem{errors} 
The errors for $\phi_2$ seem considerable, 
to make a small but non-vanishing coefficient of 
$t^{-3/2}$ in the asymptotic expansion of $\phi_2 (x,t)$, 
unsuitably for the expected behavior (\ref{eqn:380}). 
However, 
$\phi_2 (x,t)$ may still follow Eq.\ (\ref{eqn:380}) at long times, 
until the term proportional to $t^{-3/2}$ dominates.
\bibitem{situation} 
One may notice that the situation considered here 
is just the same as in Fig.\ 18 in the textbook \cite{Schiff}.
\end{thebibliography}
\end{document}